\documentclass{article}
\usepackage{comment}
\usepackage{graphicx}
\usepackage{subfigure}
\usepackage{xcolor}
\usepackage{geometry}
\usepackage{caption}
\usepackage{amsmath}
\usepackage{hyperref}
\usepackage{algorithm}
\usepackage{algpseudocode}
\usepackage{algorithmicx}

\newtheorem{theorem}{Theorem}
\newtheorem{obs}{Observation}
\usepackage{multirow}
\title{Semi-dynamic Algorithms for Strongly Chordal Graphs}
\author{Md. Zamilur Rahman and Asish Mukhopadhyay \\
	School of Computer Science \\
	University of Windsor}

\begin{document}
\maketitle

%\chapter[Semi-dynamic Algorithms for Strongly Chordal Graphs]{Semi-dynamic Algorithms for Strongly Chordal Graphs}
%\chaptermark{Strongly Chordal Graph Generation}
%\label{ch_Dynamic_Strongly_Chordal_Graph}

\begin{abstract}
There is an extensive literature on dynamic algorithms for a large number of graph theoretic problems, particularly 
for all varieties of shortest path problems. Germane to this paper are a number fully dynamic algorithms 
that are known for chordal graphs. However, to the best of our knowledge
no study has been done for the problem of dynamic algorithms for strongly chordal graphs. 
To address this gap, in this paper, we propose a  semi-dynamic algorithm for edge-deletions and a semi-dynamic
algorithm for edge-insertions in a strongly chordal graph, $G = (V, E)$, on $n$ vertices and $m$ edges. 
The query complexity of an edge-deletion is $O(d_u^2d_v^2 (n + m))$, where $d_u$ and $d_v$ are the degrees of the 
vertices $u$ and $v$ of the candidate edge $\{u, v\}$, while the query-complexity of an edge-insertion is $O(n^2)$. 
\end{abstract}

\section{Introduction}
Let $G = (V, E)$ be a connected graph on $n (=|V|)$ vertices with $m (= |E|)$ edges. $G$ is chordal if it has no induced chordless cycle of size greater than three. An alternate characterization that is more apposite for this paper is there exists 
a perfect elimination ordering of its vertices. We will represent a chordal graph by a clique tree 
(see~\cite{10.1007/978-1-4613-8369-7_1}). 

Strongly chordal graphs which are a subclass of chordal graphs was introduced into the graph theory literature by Farber in 
\cite{FarberThesis}. $G$ is strongly chordal if there exists a strong elimination ordering of its vertices, 
a generalization of the notion of perfect elimination ordering for chordal graphs. 

While the problems of finding a minimum weight dominating set and an independent dominating set in
vertex-weighted chordal graphs are NP-hard, Farber again~\cite{DBLP:journals/dam/Farber84} showed that these can be solved 
in linear time for strongly chordal graphs, provided a strong elimination ordering of the vertices of the graph is known. 

There is an extensive literature on dynamic algorithms for a large number of graph theoretic problems, particularly 
for all varieties of shortest path problems. Germane to this paper are a number fully dynamic algorithms 
that are known for chordal graphs ~\cite{DBLP:journals/talg/Ibarra08},~\cite{DBLP:journals/tcs/Mezzini12}. However, to the best of our knowledge
no study has been done for the problem of dynamic algorithms for strongly chordal graphs. 
 
To address this gap, in this paper, we propose a  semi-dynamic algorithm for edge-deletions and a semi-dynamic
algorithm for edge-insertions in a strongly chordal graph. 

\begin{comment}
The proposed semi-dynamic algorithms update strongly chordal graphs after deletion or insertion of an edge. To delete an edge from a strongly chordal graph, at first we check the deletion of an edge to preserve the chordality condition and then we check the deletion of the edge to preserve the strong chordality condition. We then delete the edge after satisfying both the conditions and update both the clique tree and the graph. On the other hand, the semi-dynamic algorithm under the insertion of an edge, we generate a strong elimination ordering from the given strongly chordal graph and then create a neighborhood matrix. If the insertion of an edge does not generate any submatrix $\bigl[ \begin{smallmatrix}1 & 1\\ 1 & 0\end{smallmatrix}\bigr]$ in the neighborhood matrix, then we insert the edge and update the graph.
\end{comment}

The rest of the paper is structured as follows. In the next section, we explain the design of semi-dynamic algorithm for 
deletions. In the following section~\ref{sec_Semi-dynamic_Algorithm_for_Insertions}, we discuss the complexity of the proposed semi-dynamic algorithms. Finally, section~\ref{sec_Dynamic_Strongly_Chordal_Discussion} contains concluding remarks and open problems.

\section[Semi-dynamic Algorithm for Deletions]{Semi-dynamic Algorithm for Deletions}
\sectionmark{Semi-dynamic Algorithm for Deletions}
\label{sec_Semi-dynamic_Algorithms_for_Deletions}

\begin{comment}
This section describes two semi-dynamic methods for the maintenance of strongly chordal graphs under the deletion and insertion of an edge. For a given strongly chordal graph ($G$), both methods take an edge $\{u,v\}$ as input to delete or insert. The following subsections explain both methods with algorithms and examples.
\end{comment}

%\subsection{Deletion of an Edge from a Strongly Chordal Graph}

Let $G = (V, E)$ be a strongly chordal graph and $C$ an even cycle of size six or 
greater in $G$. A chord $\{u, v\}$ of $C$ is a strong chord if a shortest distance between $u$ and $v$ along $C$, $d_C(u, v)$, is odd. 
The deletion algorithm is based on the following characterization of a strongly chordal graph. 

\begin{theorem}{\rm~\cite{DBLP:journals/dm/Farber83}}
	A graph $G$ is strongly chordal if and only if it is chordal and every even cycle of length at least 6 in $G$ has a strong chord.
\end{theorem}

Let $e = \{u, v\}$ be an arbitrary edge of $G$. 
Then $e$ can be deleted from $G$ provided $G-e$ remains chordal and it is not the only strong chord of a six-cycle. The check for chordality exploits the following theorem. 

\begin{theorem}{\rm~\cite{DBLP:journals/talg/Ibarra08}}
	Let $e$ be an edge of a chordal graph $G$. Then $G-e$ remains chordal if and only if $G$ has exactly one maximal clique containing $e$.
\end{theorem}

Since strongly chordal graphs are a subclass of chordal graphs, a clique tree data structure, $T$, representing $G$ is used to check for the chordality condition. 

%A clique tree is defined as follows:
%
%%http://www.cs.toronto.edu/~stacho/public/splitm1.pdf
%%plus Ibarra paper
%\begin{definition}
%	Let $G$ be a connected chordal graph and $T$ be any clique tree of $G$. The nodes of $T$ are the maximal cliques of $G$ such that for every two maximal cliques $K$, $K'$, each clique on the path from $K$ to $K'$ in $T$ contains $K\cap K'$. The edges of $T$ are labeled as $|K\cap K'|$.
%\end{definition}
%Thus, any $T$ of $G$ has nodes and edges that respectively correspond to the maximal cliques and minimal vertex separators of $G$.

\begin{figure}[htb]
	\centering
	\subfigure[A chordal graph ($G$)\label{Fig-SCGExample}]{\includegraphics[scale=.65]{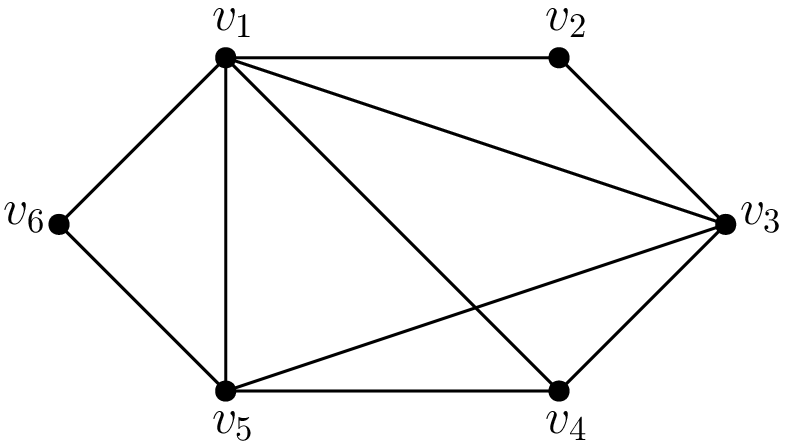}}\hspace{1in}
	\subfigure[A clique tree ($T$) \label{Fig-SCGExample-2}]{\includegraphics[scale=.65]{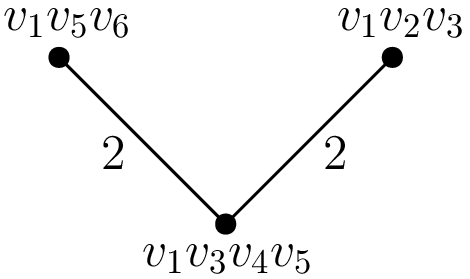}}
	\caption{An example}%
	\label{Fig:Examples}%
\end{figure}

Consider the chordal graph shown in Figure~\ref{Fig-SCGExample} that has three maximal cliques $v_1v_2v_3$, $v_1v_3v_4v_5$, and $v_1v_5v_6$. Each maximal clique is represented by a node in the clique tree $T$ and the weight of each edge is the size of the overlap of the two maximal cliques that it joins. To obtain a clique tree $T$ from $G$, we use an expanded version of the Maximum Cardinality Search (MCS) algorithm by Blair and Peyton~\cite{BlairPeyton1993}. % in $O(|V|+|E|)$ time.
If an edge $e$ is present in two or more clique tree nodes, then we cannot delete $e$ as its deletion will violate the chordality property of $G$. For instance, we cannot delete the edge $\{v_1,v_5\}$ or the edge $\{v_1,v_3\}$ from $G$ (see Figure~\ref{Fig:Examples}) because with the deletion of either of these two edges $G$ will cease to be chordal. Thus by maintaining the clique tree data structure $T$, we can determine if an edge can be deleted without violating chordality. This is done as follows. For each node in $T$, we compute the intersection of the node (a maximal clique contains two or more vertices of $G$) with the edge $e$. If we find $T$ has exactly one node containing the end-points of $e$, then we continue and check if the deletion preserves strong chordality. 

\begin{figure}[htb]
	\centering
	\includegraphics[scale= 0.65]{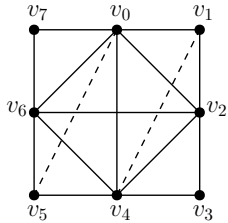}
	\caption{{\em A strongly chordal graph}}
	\label{Fig-TrampolineStronglyChordalGraph}
\end{figure}

As explained earlier, an edge $e$ can be deleted if and only it is not the only strong chord of a six-cycle. For instance, consider the strongly chordal graph shown in Figure~\ref{Fig-TrampolineStronglyChordalGraph}. The edge $\{v_0,v_5\}$ (shown as a dashed line segment) splits the 8-cycle, $\langle v_0, v_1, v_2, v_3, v_4, v_5, v_6, v_7, v_0 \rangle$, into a 4-cycle, 
$\langle v_0, v_5, v_6, v_7, v_0 \rangle$, and a 6-cycle, $\langle v_0, v_1, v_2, v_3, v_4, v_5, v_0 \rangle$. The addition of a strong chord $\{v_1,v_4\}$ (shown as a dashed line segment) splits the 6-cycle, $\langle v_0, v_5, v_4, v_3, v_2, v_1, v_0 \rangle$, into two 4-cycles, $\langle v_0, v_1, v_4, v_5, v_0 \rangle$ and $\langle v_1, v_2, v_3, v_4, v_1 \rangle$. Alternately, we could interpret $\{v_0,v_5\}$  as a strong chord of the 6-cycle $\langle v_0, v_7, v_6, v_5, v_4, v_1, v_0 \rangle$, post the introduction of $\{v_1,v_4\}$ as a strong chord of the initial 8-cycle. 

Since the addition of a strong chord splits an even cycle of size 6 or greater into two odd-length paths, it is sufficient to check for the presence of a strong chord in every even cycle of length 6. We give a formal proof of this in an Appendix. 

A potential 6-cycle of which $e =\{u, v\}$ is a strong chord is formed by disjoint pairs of chordless $P_4$ paths (each spanning four vertices) that 
go from $u$ to $v$. Thus we determine all such $P_4$ paths and for every disjoint pair of these we check whether $\{u,v\}$ is the only strong chord or not. If there is a strong chord other than $\{u,v\}$ in every disjoint pair of $P_4$ paths, then the edge $\{u,v\}$ can be deleted. On the other hand, if $\{u,v\}$ is the only strong chord for any disjoint pair of $P_4$ paths, then the edge $\{u,v\}$ cannot be deleted. 

Now, instead of searching for $P_4$ paths in the entire graph, we do a local search in an induced graph, called {\em AuxGraph}. This graph is induced by a  set of vertices {\em AuxNodes} that includes the closed neighbors of both the end-points of the edge $\{u,v\}$ to be deleted. Formally, we define $AuxNodes = N[u]\cup N[v]$. 

Setting $u$ as the source vertex, we now perform breadth-first search in $G[AuxNodes]$ to find all $P_4$ paths from $u$ to $v$.

\begin{algorithm}[htb]
	\caption{Delete}\label{algoscgdel}
	\begin{algorithmic}[1]
		\Require A strongly chordal graph $G$ and an edge $\{u,v\}$ to be deleted
		\Ensure A strongly chordal graph $G - \{u,v\}$
		\If {$Delete-Query(G,T,u,v)$ returns ``True"}
			\State Call UpdateCliqueTreeAfterDeletion
			\State Delete the edge $\{u,v\}$ from $G$
		\EndIf
	\end{algorithmic}
\end{algorithm}
\begin{algorithm}[ht]
	\caption{Delete-Query}\label{algoscgdelque}
	\begin{algorithmic}[1]
		\Require A strongly chordal graph $G$, a clique tree $T$ of $G$, and an edge $\{u,v\}$ to be deleted
		\Ensure Return True or False
		\State $canBeDeleted\leftarrow$ False
		\If {the edge $\{u,v\}$ does not exist}
		\State $canBeDeleted\leftarrow$ False
		\State \Return $canBeDeleted$
		\Else
		\If {the edge $\{u,v\}$ belongs to exactly one node ($x$) in $T$}
		\If {$\{u, v\}$ is not a strong chord}
		\State $canBeDeleted\leftarrow$ True
		\State \Return $canBeDeleted$
		\Else 
		\State $canBeDeleted\leftarrow$ False
		\State \Return $canBeDeleted$
		\EndIf
		\Else 
		\State $canBeDeleted\leftarrow$ False
		\State \Return $canBeDeleted$
		\EndIf
		\EndIf
	\end{algorithmic}
\end{algorithm}
Algorithm~\ref{algoscgdelque} returns ``True", if $\{u,v\}$ can be deleted from $G$. When the algorithm returns ``True", we perform the delete operation (see Algorithm~\ref{algoscgdel}). After performing the delete operation, we update both the clique tree and the graph. The clique tree node that contains the edge $\{u, v\}$ can be replaced with 0, 1, or 2 nodes. The algorithm $UpdateCliqueTreeAfterDeletion$ due to Ibarra~\cite{DBLP:journals/talg/Ibarra08} deletes the edge $\{u,v\}$ and updates $T$.

%\begin{comment}
\begin{algorithm}[t]
	\caption{UpdateCliqueTreeAfterDeletion~\cite{DBLP:journals/talg/Ibarra08}}\label{algoscgupdatect}
	\begin{algorithmic}[1]
% 		\Procedure{Distance\_Matrix\_Completion}{$R$}
		\Require A clique tree $T$ and an edge $\{u,v\}$ to be deleted
		\Ensure An updated clique tree $T$
		\State For every $y\in N(x)$, test whether $u\in K_y$ or $v\in K_y$ and whether $w(x,y)=k-1$ \\
		\Comment{$K_x$ and $K_y$ are maximal cliques and $w(x,y) = |K_x \cap K_y|$} 
		\State Replace node $x$ with new nodes $x_1$ and $x_2$ respectively representing $K^u_x$ and $K^v_x$ and add edge $\{x_1, x_2\}$ with $w(x_1, x_2) = k-2$.\\
		\Comment{$K^u_x = K_x - \{v\}$ and $K^v_x = K_x - \{u\}$}
		\If {$y\in N_u$}
		\State replace $\{x,y\}$ with $\{x_1, y\}$
		\EndIf
		\If {$z\in N_v$}
		\State replace $\{x, z\}$ with $\{x_2, z\}$
		\EndIf
		\If {$w\in N_{\overline{uv}}$}
		\State replace $\{x,w\}$ with $\{x_1,w\}$ or $\{x_2,w\}$ (chosen arbitrarily)
		\EndIf
		\If {$K^u_x$ and $K^v_x$ are both maximal in $G-\{u, v\}$}
		\State stop
		\EndIf
		\If {$K^u_x$ is not maximal because $K^u_x\subset K_{y_i}$ for some $y_i\in N_u$}\label{kuxS}
		\State choose one such $y_i$ arbitrarily, contract $\{x_1, y_i\}$, and replace $x_1$ with $y_i$
		\EndIf\label{kuxE}
		\If {$K^v_x$ is not maximal}
		\State perform similar operations as step~\ref{kuxS} to~\ref{kuxE}
		\EndIf
		\State \Return the clique tree $T$
% 		\EndProcedure
	\end{algorithmic}
\end{algorithm}
%\end{comment}

\begin{figure}[htb]
	\centering
	\subfigure[A strongly chordal graph ($G$)\label{Fig-SCGExampleSec}]{\includegraphics[scale=.65]{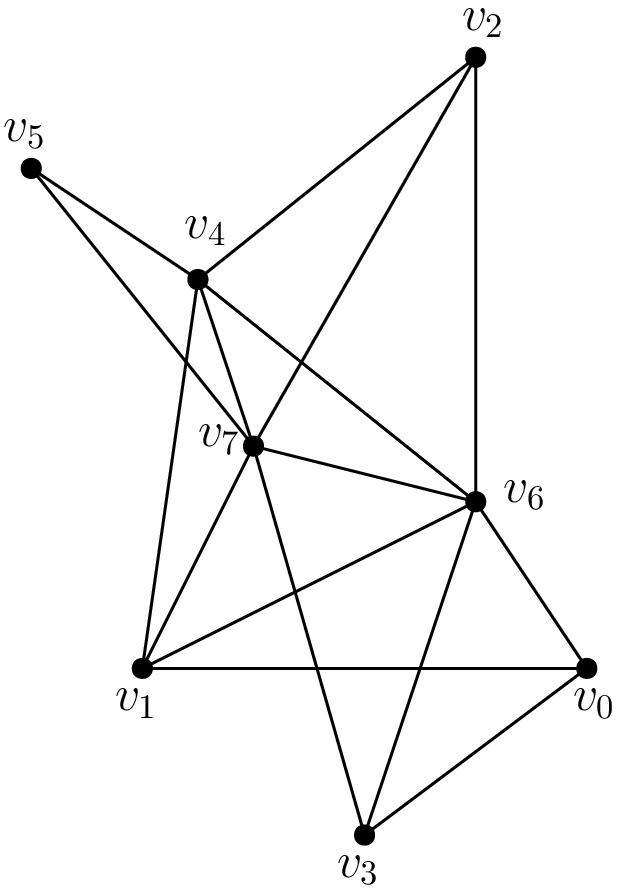}}\hspace{1in}
	\subfigure[A clique tree ($T$) of $G$ \label{Fig-SCGExample-5}]{\includegraphics[scale=.65]{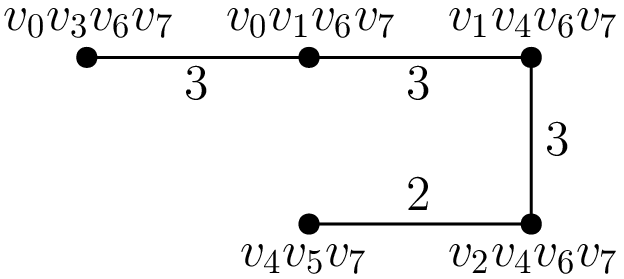}}\hspace{1in}
	\subfigure[$T-\{v_4,v_5\}$ \label{Fig-SCGExample-6}]{\includegraphics[scale=.65]{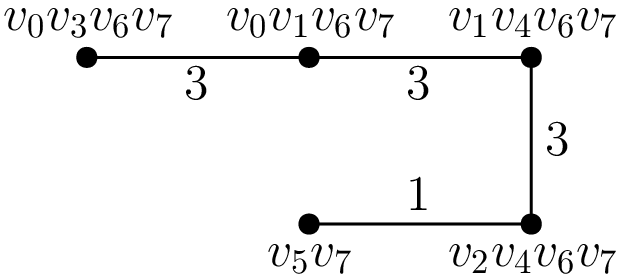}}\hspace{1in}
	\subfigure[$G-\{v_4,v_5\}$ \label{Fig-SCGExample-7}]{\includegraphics[scale=.65]{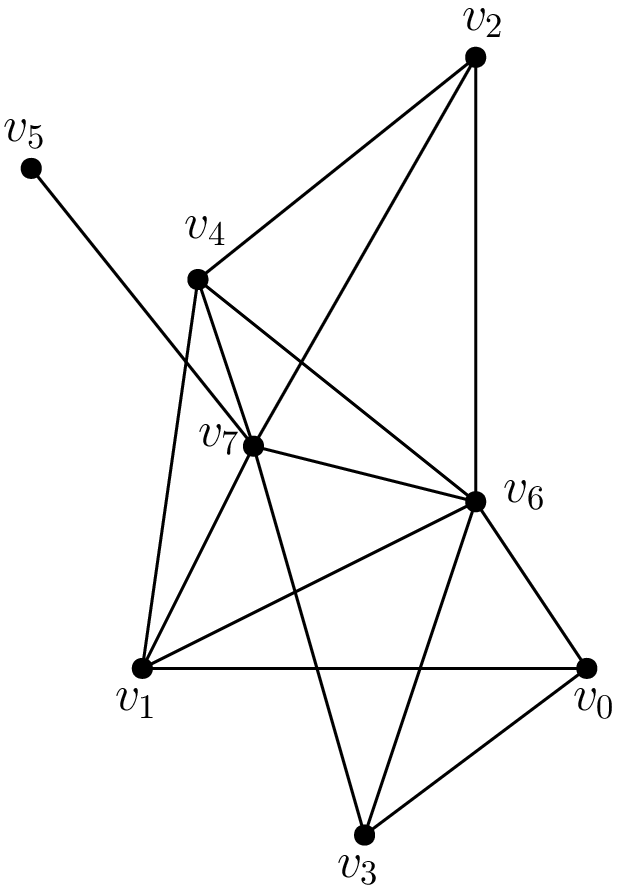}}
	\caption{An example}%
	\label{Fig:Examples}%
\end{figure}
Consider the strongly chordal graph $G$ shown in Figure~\ref{Fig-SCGExample}. Can we delete the edge $\{v_1,v_4\}$? At first, we check the chordality property. From the clique tree $T$ (see Figure~\ref{Fig-SCGExample-2}), we observe there is a single node containing the edge $\{v_1,v_4\}$ only. This satisfies the chordality condition and now we check the strong chordality condition. To do that, we compute the closed neighborhood of $v_1$ and $v_4$ and then compute the {\em AuxNodes} set, where $AuxNodes = N[v_1]\cup N[v_4] = \{v_1, v_2, v_3, v_4, v_5, v_6\}\cup \{v_4, v_1, v_3, v_5\}=\{v_1, v_4, v_2, v_3, v_5, v_6\}$. There are two $P_4$ paths ($v_1-v_2-v_3-v_4$ and $v_1-v_6-v_5-v_4$) between $v_1$ and $v_4$ in the {\em AuxGraph} created from the {\em AuxNodes} set. We notice that $\{v_1,v_4\}$ is the only strong chord in the {\em AuxGraph}. Hence the deletion of $\{v_1,v_4\}$ is not allowed. But in the presence of any of the other two strong chords ($\{v_2,v_5\}$ or $\{v_3,v_6\}$) in the {\em AuxGraph} we will be able to delete $\{v_1,v_4\}$. 

Consider another example graph, shown in Figure~\ref{Fig-SCGExampleSec} where the chordality is preserved if we delete $\{v_7,v_2\}$, but the edge $\{v_7,v_2\}$ is the only strong chord for the 6-cycle ($v_7-v_1-v_6-v_2-v_5-v_4-v_7$). So, the edge $\{v_7,v_2\}$ cannot be deleted. But the edge $\{v_5,v_4\}$ can be deleted from the graph. After checking the clique tree, we see that only a single node contains the edge $\{v_5,v_4\}$. So we can proceed to check for the strong chordality condition. We observe that $\{v_5,v_4\}$ is not a strong chord and thus we can delete the edge $\{v_5,v_4\}$ and update both $T$ and $G$ (as shown in Figure~\ref{Fig-SCGExample-6} and~\ref{Fig-SCGExample-7}, respectively.)

\subsection{Complexity of Deletions}
\begin{figure}[t!]
	\centering
	\includegraphics[scale=0.65]{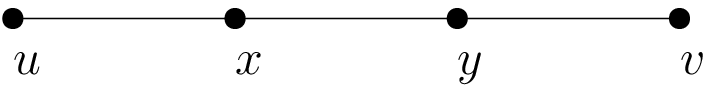}	
	\caption{A $P_4$ path from $u$ to $v$}
	\label{Fig-P4Path}
\end{figure}

For the semi-dynamic deletion, we create a clique tree $T$ in $O(|V|+|E|)$ time, using the expanded version of the MCS algorithm. To check if the edge $\{u,v\}$ belongs to exactly one maximal clique in $T$, we perform the intersection operation between a maximal clique and the edge $\{u,v\}$. This operation takes linear time. To bound the query complexity of deleting an edge $\{u, v\}$ from the strongly chordal graph, we note that this is dominated by the case of finding multiple $P_4$ paths between $u$ and $v$ and we have to consider these in pairs and run the breadth-first search. 
%The number of pairs $P_4$'s can be estimated in the same way as did for strongly chordal graphs in %section~\ref{subsec_weaklyComplexity}, where we have an upper bound on path count is $O(d_u^2d_v^2)$. 
An upper bound on the number of pairs of $P_4$ paths between $u$ and $v$ is $O(d_u^2d_v^2)$, where $d_u$ and $d_v$ are 
the degrees of $u$ and $v$ respectively. For consider such a path 
from $u$ to $v$ (Fig.~\ref{Fig-P4Path}): $x$ is one of the at most $d_u$ vertices adjacent to $u$ and $y$ is one of the at most $d_v$ vertices adjacent to $v$, so that we have at most $O(d_ud_v)$ $P_4$ paths from $u$ to $v$ and thus $O(d_u^2d_v^2)$ 
disjoint pairs of $P_4$ paths from $u$ to $v$. 

If $|E|$ is the number of edges currently, in the strongly chordal graph, the complexity of running a breadth-first-search is $O(n + |E|)$. Since $m$ is the number of edges in the final strongly chordal graph, an upper bound on the query complexity is $O(d_u^2d_v^2 (n + m))$.

The deletion of an edge take constant time since we maintain an adjacency matrix data structure to represent $G$. This is in addition to the clique tree that we maintain. It is however possible to dispense with the adjacency matrix. In that case, there is 
additional work required to construct the graph $G$ from the clique tree. This can be done using a certain Running Intersection 
Property of chordal graphs (see \cite{BlairPeyton1993} for details). 

%\subsection{Insertion of an Edge into a Strongly Chordal Graph}
\section{Semi-dynamic Algorithm for Insertions}\label{sec_Semi-dynamic_Algorithm_for_Insertions}

Let $\alpha = \langle v_1, v_2, \ldots, v_n \rangle$ be a strong elimination ordering of the vertices $V$ of a strongly chordal
graph $G$. The neighborhood matrix $M(G)$ of $G$, based on $\alpha$, is an $n \times n$ matrix whose $(i, j)$-th entry is $1$ if $v_i\in N[v_j]$ and is $0$ otherwise. Let $\Delta$ be the submatrix:

\begin{equation*}
\Delta =
\begin{bmatrix}
1 & 1 \\
1 & 0
\end{bmatrix}
\end{equation*}

Our dynamic insertion algorithm is based on the following observation.

\begin{obs}~\cite{DBLP:journals/dm/Farber83}
The row (and column) labels of $M(G)$ correspond to a strong elimination ordering if and only if the matrix $M$ does not contain $\Delta$ as a submtraix.
\end{obs}

The absence of $\Delta$ in $M$ implies that $M$ is totally balanced and the theorem below allows us to claim that 
$G$ is strongly chordal. 

\begin{theorem}~\cite{DBLP:journals/dm/Farber83}
	A graph $G$ is strongly chordal if and only if $M(G)$ is totally balanced.
\end{theorem}

Thus if $G$ is strongly chordal, then $G+\{u,v\}$ remains so if inserting the  edge $\{u,v\}$ into $G$  does not create any $\Delta$ submatrix in $M(G +\{u,v\})$.

To insert an edge $\{u,v\}$ into $G$, we first check if it is already present in $G$. If not, we insert it if no submatrix $\Delta$ is created. 

The initialization process consists of computing a strong elimination ordering $\alpha$ of an input strongly chordal graph $G$, using a recognition algorithm for strongly chordal graphs due to Farber~\cite{DBLP:journals/dm/Farber83}.  
 
\begin{algorithm}[htb]
	\caption{Insert}\label{algoscgins}
	\begin{algorithmic}[1]
		\Require A strongly chordal graph $G$ and an edge $\{u,v\}$ to be inserted
		%\Ensure A strongly chordal graph $G+\{u,v\}$
		\Ensure  Neighborhood Matrix of $G$
		\If {$Insert-Query(G,u,v)$ returns ``True"}
		\State insert edge $\{u,v\}$ into $G$
		\State {\bf return} updated $M(G)$ \Comment{$M(G)$ is the neighborhoood matrix}
		\EndIf
	\end{algorithmic}
\end{algorithm}

\begin{figure}[h]
	\centering
	\includegraphics[scale = .65]{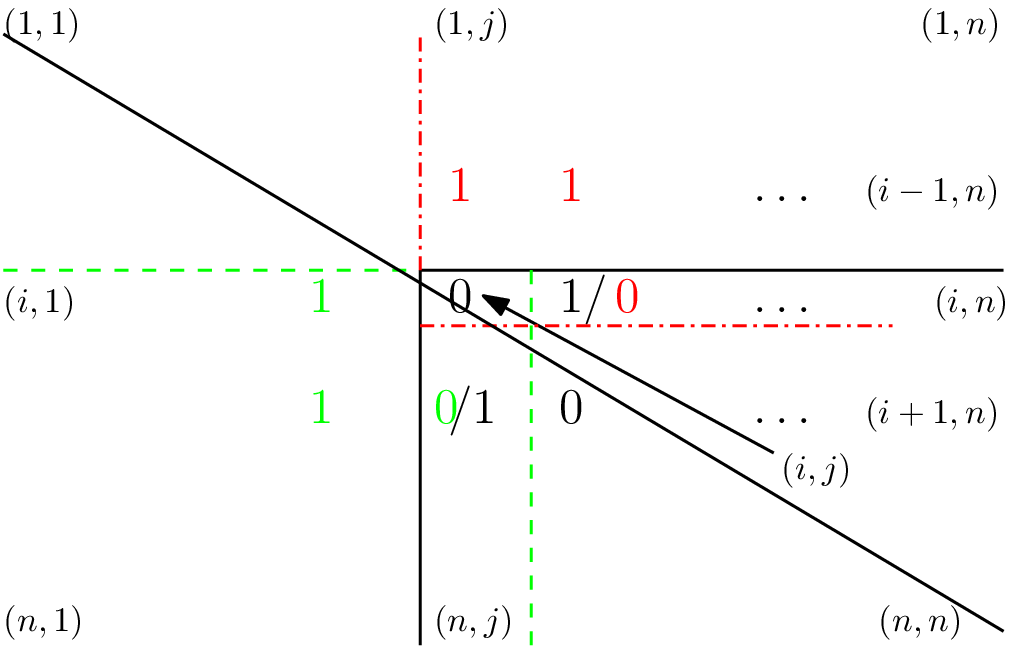}
	\caption{{\em Algorithm to find $\Delta$ submatrix}}
	\label{Fig-MatrixInsertion}
\end{figure}

After creating a strong elimination ordering $\alpha$, we create the neighborhood matrix $M(G)$ of $G$. We also find the order of $u$ and $v$ in the ordering $\alpha$. Now we check if the insertion of an edge $\{u,v\}$ creates any $\Delta$ submatrix or not. The searching strategy is explained with the Figure~\ref{Fig-MatrixInsertion}. Assume there is a 0 in the $i^{th}$ row and $j^{th}$ column. Now we want to change the entry from 0 to 1 (which corresponds to the insertion of an edge in the graph). Before changing the entry from 0 to 1, we need to check if that creates any $\Delta$ submatrix in the $M(G)$ or not. To check the presence of $\Delta$ submatrix in $M(G)$, we need to check in three different directions from the $(i,j)$-th position. We need to check in the upward direction from $(i,j)$ to $(1,n)$, in the downward direction from $(i,j)$ to $(n,n)$, and on the left direction from $(i,j)$ to $(n,1)$ position.  If there is no $\Delta$ submatrix created in $M(G)$ for changing the entry from 0 to 1, then we change the $(i,j)$-th entry from 0 to 1. Since $M(G)$ is a symmetric matrix, we also change the $(j,i)$-th entry from 0 to 1.

\begin{algorithm}[!t]
	\caption{Insert-Query}\label{algoscginsque}
	\begin{algorithmic}[1]
		\Require A strongly chordal graph $G$ and an edge $\{u,v\}$ to be inserted
		\Ensure Return True or False
		\State $canBeInserted\leftarrow$ True
		\If {$\{u, v\}$ is an edge of $G$}
		\State $canBeInserted\leftarrow$ False
		\State \Return $canBeInserted$
		\Else
		%		\State Create a strong elimination ordering (SEO) of $G$
		%		\State $i\leftarrow SEO[u]$
		%		\State $j\leftarrow SEO[v]$
		%		\State Create a neighborhood matrix $M(G)$
		\For{$l \gets j$ to $n$}\Comment downward
		\For{$k \gets i$ to $n$}
		\If {$(M[i][l+1] == 1$ and $M[k+1][j] == 1$ and $M[k+1][l+1] == 0)$}
		\State $canBeInserted\leftarrow$ False
		\State \Return $canBeInserted$
		\EndIf
		\EndFor
		\EndFor
		
		\For{$l \gets j$ to $n$} \Comment upward
		\For{$k \gets i$ to $0$}
		\If {$(M[k-1][j] == 1$ and $M[k-1][l+1] == 1$ and $M[i][l+1] == 0)$}
		\State $canBeInserted\leftarrow$ False
		\State \Return $canBeInserted$
		\EndIf
		\EndFor
		\EndFor
		
		\For{$l \gets j$ to $0$} \Comment leftward
		\For{$k \gets i$ to $n$}
		\If {$(M[i][l-1] == 1$ and $M[k+1][l-1] == 1$ and $M[k+1][j] == 0)$}
		\State $canBeInserted\leftarrow$ False
		\State \Return $canBeInserted$
		\EndIf
		\EndFor
		\EndFor
		\State \Return $canBeInserted$
		\EndIf
	\end{algorithmic}
\end{algorithm}

\begin{figure}[htb]
	\centering
	\subfigure[A strongly chordal graph $G$\label{Fig-SCGExampleIns}]{\includegraphics[scale=.65]{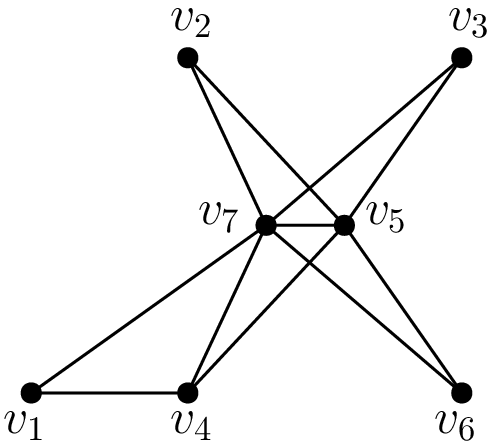}}\hspace{50pt}
	\subfigure[$G+\{v_1,v_6\}$\label{Fig-SCGExampleIns-1}]{\includegraphics[scale=.65]{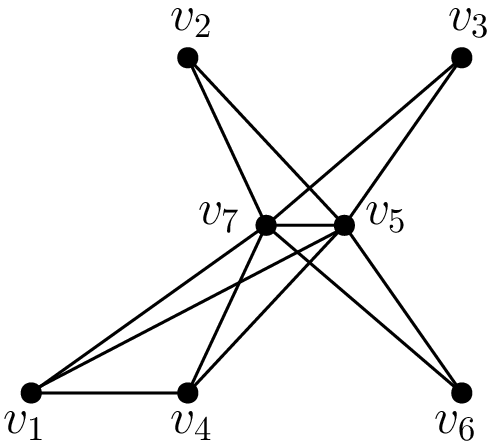}}\\
	\subfigure[A neighborhood matrix $M(G)$ of the strongly chordal graph shown in Figure (a)]
	{$\begin{bmatrix}
%		v_1 & v_2 & v_3 & v_4 & v_6 & v_5 & v_0 \\
		1   & 0   & 0   & 1   & 0   & 0   & 1   \\
		0   & 1   & 0   & 0   & 1   & 0   & 1   \\
		0   & 0   & 1   & 0   & 1   & 0   & 1   \\
		1   & 0   & 0   & 1   & 1   & 0   & 1   \\
		0   & 1   & 1   & 1   & 1   & 1   & 1   \\
		0   & 0   & 0   & 0   & 1   & 1   & 1   \\
		1   & 1   & 1   & 1   & 1   & 1   & 1
	\end{bmatrix}$}\hfil
	\subfigure[A neighborhood matrix $M(G)$ of the strongly chordal graph shown in Figure (b)]
	{$\begin{bmatrix}
%		v_1                 & v_2 & v_3 & v_4 & v_6                 & v_5 & v_0 \\
		1                   & 0   & 0   & 1   & \textbf{\textit{1}} & 0   & 1   \\
		0                   & 1   & 0   & 0   & 1                   & 0   & 1   \\
		0                   & 0   & 1   & 0   & 1                   & 0   & 1   \\
		1                   & 0   & 0   & 1   & 1                   & 0   & 1   \\
		\textbf{\textit{1}} & 1   & 1   & 1   & 1                   & 1   & 1   \\
		0                   & 0   & 0   & 0   & 1                   & 1   & 1   \\
		1                   & 1   & 1   & 1   & 1                   & 1   & 1
	\end{bmatrix}$}
	\caption{An example of insertion}\label{Fig-ExamplesIns}
\end{figure}

Algorithm~\ref{algoscginsque} returns ``True" if $\{u,v\}$ can be inserted into $G$. In that case, we perform the insert operation (see algorithm~\ref{algoscgins}). Consider the strongly chordal graph shown in Figure~\ref{Fig-SCGExampleIns}. After finding a strong elimination ordering $\langle v_1, v_2, v_3, v_4, v_5, v_6, v_7 \rangle$ of $G$, we create the neighborhood matrix $M(G)$. Now, suppose we want to insert an edge, say $\{v_1,v_6\}$, into $G$. However, since the insertion of $\{v_1,v_6\}$ creates a $\Delta$ submatrix in $G$, this cannot be done. Next, we might want to insert, say edge $\{v_1,v_5\}$, into $G$. This is possible as its insertion does not create any $\Delta$ submatrix in $G$. Since $M(G)$ is a symmetric matrix, we changed 0 to 1 in both symmetric positions (which corresponds to the insertion of $\{v_1,v_5\}$ into $G$).

\subsection[Complexity]{Complexity of Insertions}
\sectionmark{Complexity}
\label{sec_Dynamic_Strongly_Chordal_Complexity}

Computing a strong elimination ordering using Farber's algorithm takes $O(n^3)$ time, and 
it takes $O(n^2)$ time to initialize the neighborhood matrix $M(G)$. Thus the preprocessing time-complexity is $O(n^3)$. 
The upper bound on searching for a $\Delta$ submatrix in $M(G)$ is $O(n^2)$. Thus the time-complexity of an $insert$-$query$ 
is $O(n^2)$.

The insertion of an edge take constant time since we maintain a neighborhood matrix data structure to represent $G$.

\section[Discussion]{Discussion}
\sectionmark{Discussion}
\label{sec_Dynamic_Strongly_Chordal_Discussion}

In this paper, we have presented semi-dynamic algorithms for deletions and insertions of edges into a strongly chordal graph.
The proposed semi-dynamic algorithms are based on two different characterizations of strongly chordal graphs. The deletion algorithm is based on a strong chord characterization, while the insertion algorithm is based on a  totally balanced matrix characterization. An interesting and challenging open problem is to come up with an efficient fully dynamic algorithm for this class of graphs.

%\bibliographystyle{acm}
%\bibliography{thesis-refs}

\appendix
\section{Strong chords of even cycles of size 6 or greater}

In this appendix, we justify why it is enough to check if an edge $e$ that 
is a candidate for deletion is a strong chord of a 6-cycle. \\

{\bf Definition:} Let $C_k$ denote a cycle with $k$ edges.\\

{\bf Defintion:} An ensemble $\mathcal{E}$ of strong chords of an even cycle with $2n$ edges, $C_{2n}$, is a set of 
$n-2$ strong chords that are pairwise disjoint, except for common end-points. \\

\begin{figure}[h!]
	\centering
	\includegraphics[scale=0.65]{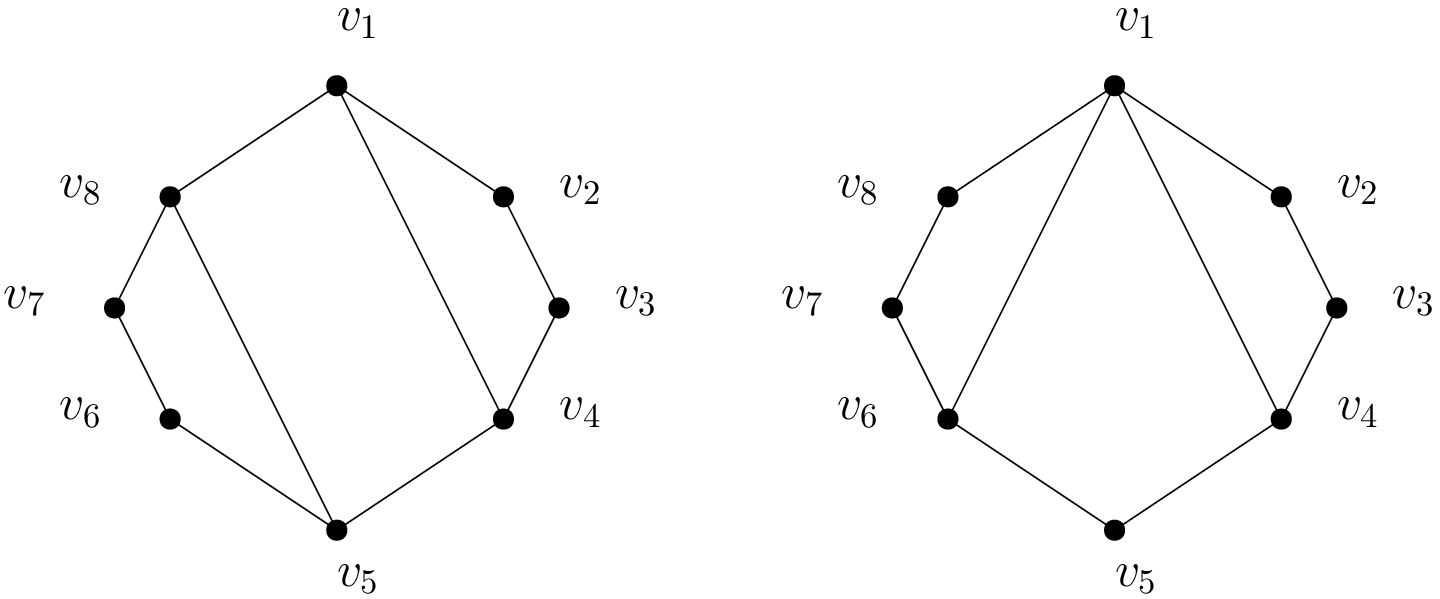}	
	\caption{Two different ensembles of strong chords of an 8-cycle}
	\label{ensemble8cycle}
\end{figure}

Two different ensembles of strong chords are shown in Fig.~\ref{ensemble8cycle} for an 8-cycle, $C_8$. 

{\bf Lemma}: A strong chord of a cycle $C_{2n}$ belongs to an ensemble of 
$n-2$ strong chords, $\mathcal{E}$. \\

{\bf Proof:} Let $v_1v_k$, for $k \geq 4$ and even, be a strong chord of an even cycle 
$C_{2n} = \langle v_1,v_2,\ldots, v_{2n}\rangle$ of 
length $2n$, where $n \geq 3$. The proof is by induction on $n$. Clearly as a single strong chord 
splits a 6-cycle into two cycles of length 4, the claim is true for $n = 3$. 
Assume that the claim is true for a cycle of length $2(n-1)$.
Since $v_1v_k$ splits $C_{2n}$ into two cycles $C_k$ and $C_{k'}$ of even lengths $k$ and $k' = 2n - k + 2$,
by the inductive hypothesis, there exists 
a disjoint ensemble of $k/2 - 2$ strong chords that partition $C_k$ and a disjoint ensemble of $(2n - k + 2)/2 - 2$ strong chords 
that partition $C_{2n-k + 2}$.  Since the strong chord $v_1v_k$ is not counted, the total number of 
strong chords that partition $C_{2n}$ is: $k/2 - 2 +(2n-k + 2 )/2 - 2 + 1 = n - 2$. This proves the assertion. \\

{\bf Theorem:} Each of the strong chords in an ensemble  $\mathcal{E}$ of strong chords of $C_{2n}$ is a strong chord of a 6-cycle. \\

\begin{figure}[t!]
	\centering
	\includegraphics[scale=0.65]{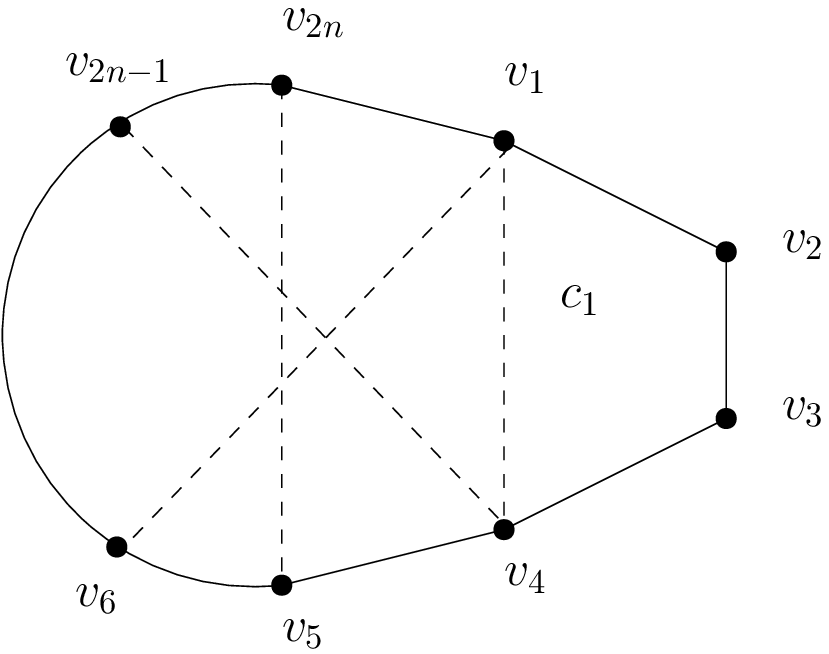}	
	\caption{Every strong chord is a strong chord of a 6-cycle}
	\label{figForStrongChordProof}
\end{figure}

{\bf Proof:} Once again the proof is by induction on $n$. Clearly this is true 
for $C_6$, as the ensemble $\mathcal{E}$  has only one strong chord. Assume the claim holds for an even cycle of smaller length. 
Among the ensemble $\mathcal{E}$ of strong chords of $C_{2n}$ there is one, say $c_1$, that forms a $C_4$ with three boundary edges (see Fig.~\ref{figForStrongChordProof}). By the inductive hypothesis, in  
the even cycle formed by the rest of the $2n-3$ edges and $c_1$ each of the strong chords of the residual ensemble $\mathcal{E} - \{c_1\}$ is a 
strong chord of a 6-cycle, $C_6$. To show that $c_1$ is also a strong chord of a 6-cycle, we observe from Fig.~\ref{figForStrongChordProof}, that
$c_1$ is a strong chord of the 6-cycle, $\langle v_{2n}v_1v_2v_3v_4v_5 \rangle$, or of the 6-cycle 
$\langle v_{2n}v_1v_2v_3v_4v_5 \rangle$ or of the 6-cycle $\langle v_{2n}v_1v_2v_3v_4v_5 \rangle$ and $v_{2n}v_5$ or $v_{1}v_6$ or
$v_{2n - 1}v_4$ is a strong chord in 
$\mathcal{E} - \{c_1\}$. This completes the proof.

\end{document}